# FIREBall-2: Flight preparation of a proven balloon payload to image the intermediate redshift circumgalactic medium

Vincent Picouet [1], David Valls-Gabaud [2], Bruno Milliard [3], David Schiminovich [1], Drew M. Miles [4], Keri Hoadley [6], Erika Hamden [7], D. Christopher Martin [4], Gillian Kyne [5], Trent Brendel [7], Aafaque Raza Khan [7], Jean Evrard [8], Zeren Lin [4], Haeun Chung [7], Simran Agarwal [7], Ignacio Cevallos Aleman [1], Charles-Antoine Chevrier [8], Jess Li [7], Nicole Melso [7], Shouleh Nikzad [5], Didier Vibert [3], Nicolas Bray [8]

[1] Columbia University, 550 W. 120 Street, New York, USA, Email: vp2376@columbia.edu
[2] Observatoire de Paris, CNRS, France, Email: david.valls-gabaud@obspm.fr
[3] Aix Marseille Université, CNRS, CNES, LAM, Marseille, France
[4] Cahill Center for Astrophysics, California Institute of Technology, Pasadena, CA 91125, USA
[5] Jet Propulsion Laboratory, California Institute of Technology, 4800 Oak Grove Drive, Pasadena, CA 91109, USA
[6] University of Iowa, Iowa City, IA 52242, USA
[7] University of Arizona, Steward Observatory, 933 N Cherry Ave, Tucson, AZ 85721, USA
[8] Centre National d'Etudes Spatiales, 31401 Toulouse Cedex 4, France

ABSTRACT

FIREBall-2 is a stratospheric balloon-borne 1-m telescope coupled to a UV multi-object slit spectrograph designed to map the faint UV emission surrounding $z\sim0.7$ galaxies and quasars through their Lyman-α line emission. This spectro-imager had its first launch on September 22$^{nd}$ 2018 out of Ft. Sumner, NM, USA. Because the balloon was punctured, the flight was abruptly interrupted. Instead of the nominal 8 hours above 32 km altitude, the instrument could only perform science acquisition for 45 minutes at this altitude. In addition, the shape of the deflated balloon, combined with a full Moon, revealed a severe off-axis scattered light path, directly into the UV science detector and about 100 times larger than expected. In preparation for the next flight, and in addition to describing FIREBall-2's upgrade, this paper discusses the exposure time calculator (ETC) that has been designed to analyze the instrument's optimal performance (explore the instrument's limitations and subtle trade-offs).

## 1. Introduction

While the depletion timescale for galaxies to convert their gas into stars is of the order of one gigayear, galaxies at all masses appear to sustain their star formation rate (SFR) for several gigayears, and this requires a supply of gas from the intergalactic medium (IGM) to fuel star formation. Therefore, studying what lies around galaxies is essential to constrain the mechanisms regulating star formation. Understanding how galaxies gather their gas could clarify why the SFR is declining since cosmic noon, whereas diffuse hydrogen is still the dominant component in galaxies [1]. The circumgalactic medium (CGM), defined as the 300 kpc interface between the galaxy and the IGM, encompasses all gas exchanges (infalls, galactic feedback, and recycling) which makes it a key ingredient to better understanding the gas reservoir supply regulation of galaxies [2]. There are numerous open questions about galaxy evolution: How do galaxies become and remain passive? How do galaxies fuel their star formation? Why do dark matter halos of different masses give rise to galaxies with drastically different star-formation rates and histories? Why does such a small fraction of cosmic baryons and metals reside within the galaxies?

All these questions can be translated into CGM-related questions, as they all feature the regulation of gas flows into and out of galaxies, which necessarily pass through the CGM:

- **Reservoir content:** What is the CGM made of? What is the total amount of gas/metals within the CGM? How is it evolving with cosmic time? Can this explain the decrease in star formation efficiency? How is the gas/energy distributed?
- **Accretion mode:** How does IGM pristine gas enter the galaxy? Does it follow cold accretion from dense flows of cold gas or hot accretion of more diffuse gas from the halo?
- **CGM / host galaxy relation:** What are the primary connections between CGM properties (content, size, etc.) and galaxy properties (dark matter and stellar mass, SFR, type, color, etc.)?

While the number and depth of observations at high redshift ($z>2$) have increased dramatically with instruments like MUSE [3] (which detects bright Lyman-alpha haloes around most galaxies at $z>3$) and KCWI, emission data are very scarce in the 10-billion-year span from $z\sim2$ to the present because of the difficulties inherent to vacuum UV observations. This is precisely the redshift gap that is explored by FIREBall-2: a CGM emission pathfinder and technology demonstrator co-funded by NASA and CNES. Despite a reduced science impact due to a balloon failure shortening the first flight in 2018 [4], this launch marked the first multi-object acquisition from space using a multi-slit spectrograph and allowed suborbital demonstration of several key technologies: δ-Doped EMCCD detector and electronics, aspherized grating, and CNES sub-arcsecond pointing system for stratospheric gondolas.

The results of the 2018 flight validated the instrument design as well as its sub-systems and predicted a gain factor of 30-40 in sensitivity for a future flight in nominal conditions. It demonstrates FIREBall-2's capability to image the CGM, the number one recommendation from Astro2020 Decadal Survey Panel on Galaxies to understand "*Cosmic Ecosystems*" over the next decade. FIREBall-2 is currently the only proven UV instrument able to make these necessary discoveries at intermediate redshift ($0.2<z<1$). In consequence, preparations for a second flight from Fort Sumner, in Fall 2023, have been performed and final corrections are currently underway. Relatively minor modifications have been performed in order to minimize change-associated risks and ensure the full efficiency gain from the last flight.

In this paper, the instrument design and in-flight performance are summarized, and the course of action to solve the different issues which appeared during the flight are presented. The predicted optimal sensitivity is analyzed thanks to the implementation of a new FIREBall-specific exposure time calculator (ETC). This ETC is also used to explore more subtle trade-offs to boost the instrument's optimal sensitivity. Further upgrades and long-term improvements are also discussed.

2. FIREBall-2 design

FIREBall-2 is designed to tackle the challenges of space and to image the intermediate-redshift CGM in emission at a low cost compared to orbital projects. The FIREBall-2 gondola has a height of ~5 meters and a weight of ~2.3 metric tons, including the 500-kg ballast. Because the atmosphere absorbs most of the UV light coming from space, the FIREBall spectrograph has been optimized to operate in the narrow atmospheric transmission window around 200 nm, where there is a dip in the atmospheric UV absorption above 35 km. The optical design of FIREBall-2 relies on a low-inertia 1.2-meter siderostat mounted on a controlled gimbal system that stabilizes the reflected beam in the gondola's frame and directs it to a fixed $f/2.5$ paraboloid that in turn focuses it at the entrance of the $f/2.5$ instrument (see Fig. 1). Target selection is achieved with a series of pre-installed precision slit mask systems. The detector is a *Teledyne-e2v* electron-multiplying CCD (EMCCD) [5] anti-reflection-coated and delta-doped by the Jet Propulsion Laboratory (JPL).

Inside the tank, the narrow field of view of the parabola alone is extended to ~800 arcmin$^2$ by a Schwarzschild-type two-mirror field corrector $f/2.5 - f/2.5$. The spectrograph itself consists of identical $f/2.5$ Schmidt collimator and camera mirrors, the aspheric correction being provided by the 2400 g/mm reflection grating obtained by double replication of a deformable matrix.

3. 2018 flight & outcomes

During the first flight, a balloon gas leakage reduced the telescope's scientific acquisition to less than an hour [6]. The 3D trajectory of the flight is presented in Fig. 2. However, the processing of the collected data made it possible to validate the instrument by obtaining the first scientific results [4]. The major operational successes of the 2018 flight are summarized below:

- first multi-object acquisition from space using a multi-slit spectrograph (any band).
- validation of the pointing control and stability of the 4-axis guidance gondola (<0.5" RMS/sec/axis, below requirement despite high excitation of the over rapid descent).
- first science flight of a high-QE UV electron-multiplying CCD (EMCCD), δ-doped and anti-reflection-coated by JPL. This enabled a first characterization of this type of detector and its electronics in an extreme, sub-orbital environment (T~150 K).
- demonstration of end-to-end instrument efficiency as expected.
- multiple-detections through the slits demonstrate the success of the complex end-to-end alignment and guidance as well as their stability.
- operation of a complete, end-to-end, highly complex/fascinating payload and a gondola which is an important process in the scientific, technical, and management training of the multiple Ph.D. candidates and postdoc involved in the project.

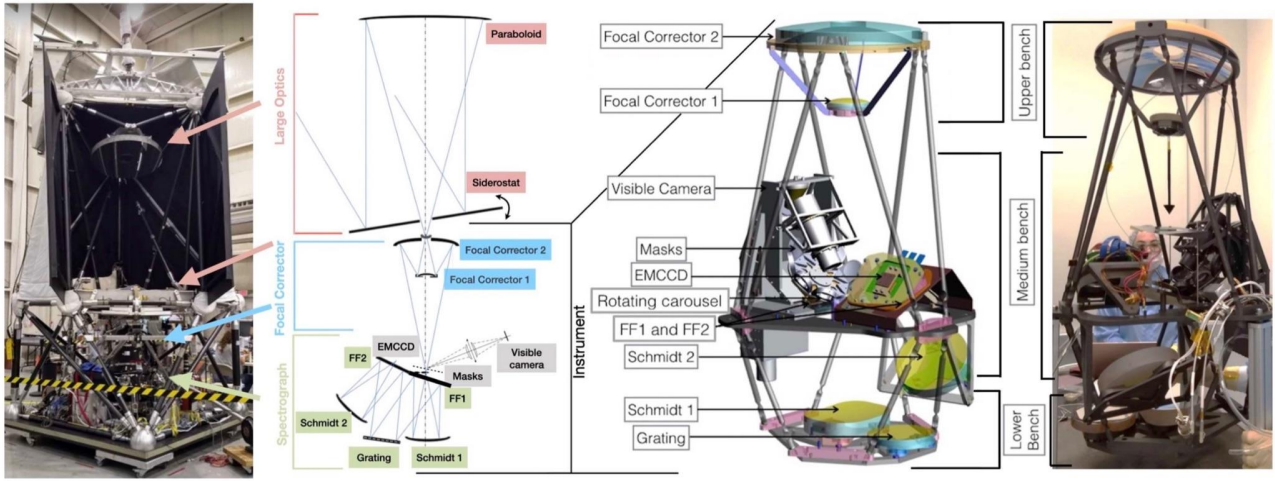

**Figure 1. Left**: FIREBall-2 picture and optical design layout. For readability, the instrument is shown in the payload layout as a flattened optical layout, rather than as a compact real-world 3-D design. **Right**: 3D design and picture of the interior of the FIREBall-2 spectrograph. Note the rotation of about 180 degrees along the z-axis compared to the left side in order to improve the visualization of the different components. During the flight, this equipment is in a vacuum tank kept at ~ $10^{-6}$ mbar by cryo-pumping and cryo-sorption at ~100 K which reduces heat losses in the detector cooling chain by minimizing ice trapping, and cools the detector near 170 K. The whole vacuum tank is placed on tip-tilt actuators to control the focus and centering during the flight.

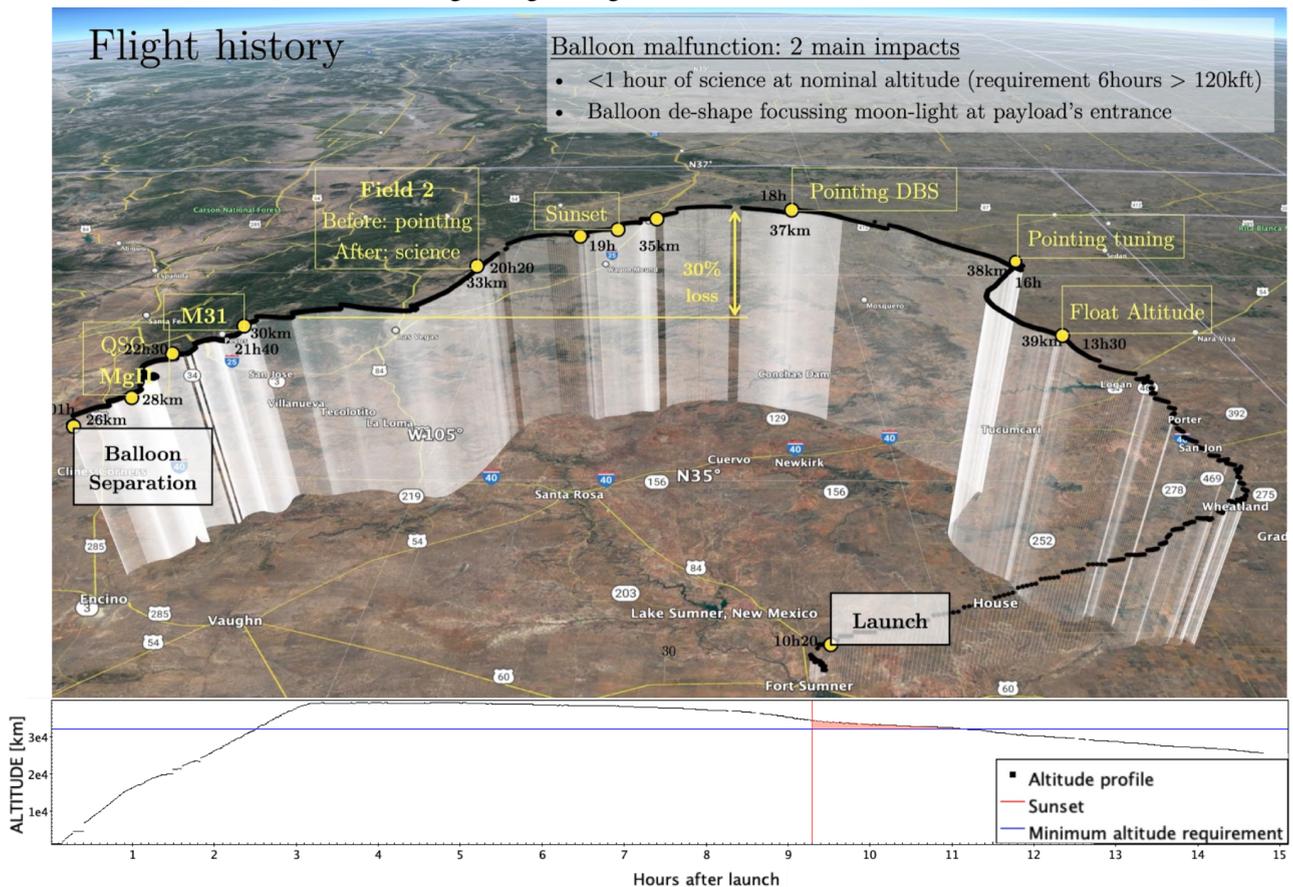

**Figure 2. Top:** FIREBall-2 2018 flight 3D trajectory (longitude, latitude, and altitude). The transparency of the vertical lines shows the instrument's ground-projected velocity. The principal flight operations are added in yellow. **Bottom:** Instrument's GPS altitude. The filled red area represents the time after sunset above the nominal altitude requirement.

The 2018 flight analysis allowed us to significantly improve the understanding of the instrument and helped the team analyze the in-flight performance and refine the noise budget of FB2. In particular, it allowed us to: (i) have a first sky background estimation at 40 km altitude, (ii) validate the instrument throughput and noise budget, (iii) assess cosmic-ray (CR) related surface loss, and (iv) analyze the charge transfer efficiency and its impact on the photon-counting thresholding process. The in-flight performance metrics are summarized in Table 1 and described in detail in [4]. The significantly degraded performance led to a limiting 5σ flux detection of $1.5\times10^{-17}$ erg/cm$^2$/s/arcsec$^2$ for a diffuse object on one element resolution and $10^{43}$ erg/s for compact objects.

Because of the flight night-time termination, the instrument experienced higher-than-expected winds upon landing, damaging key components of FB2, including the telescope mirrors (the main optics, reducing the collecting area by 10%, and the focal corrector (FC) which had to be rebonded and realigned), guidance motors, and the gondola carbon mechanical structure.

4. Preparation for the next flight

4.1. FB2 ETC and instrument's optimal sensitivity

To confirm the re-flight readiness of the FIREBall-2 instrument, these updated parameters were used to design a FB-specific ETC. This ETC aims at exploring the ultimate efficiency the instrument could reach in nominal flight conditions or after mitigation, and check its evolution under different scenarios. While this ETC is fairly simple and could be easily adapted to different projects, it includes several FB-related specificities that can have a dramatic impact on the final sensitivity (cosmic-ray-related surface loss, 40km-altitude sky-background, CTE, amplification model, optimal threshold computation and efficiency, flux cut by the slit, etc.). In addition, [post-]flight EMCCD data were analyzed in detail to understand the complexity and potential of these devices.[1] One outcome of this work was the development of an EMCCD model to analyze the detector performance. Compared to the Poisson-Gamma-Normal EMCCD likelihood model developed [5], the addition of smearing in the model allowed us to reproduce all the detector data histograms and infer automatically the associated parameters (read noise, amplification gain, smearing, serial and semi-amplified clock-induced charges (CIC), flux) making it an extremely powerful tool (see Fig. 3). This smearing-included EMCCD model is more realistic and was incorporated into the ETC to automatically compute the photon-counting threshold that optimizes the SNR based on different parameters.

The full ETC, shown in Fig. 4, and available at https://fireball-etc.lam.fr, shows the budget from the different contributors to a frame (signal, sky background, dark current, CIC, guider-related background, read noise), as well as the noise budget and the final SNR on a FB-flight-like science acquisition and the limiting flux for point-like and diffuse sources. The ETC is designed to provide direct visualization of these budgets with respect to the different flight/detector parameters that can be tuned and/or mitigated, namely:

- **Mission parameters**: acquisition time, PSF size at the mask and detector level, thresholding (or not), sky background, exposure time, readout time.
- **EMCCD parameters:** amplification gain, device temperature (determining dark current and CTE), read noise, CIC, CR rate or induced loss

This ETC GUI is extremely useful to provide at a glance not only the main instrument parameters and limitations, but also the tradeoffs, possible optimization and mitigation strategies. For example, it confirms the sensitivity limit measured on the 2018 flight data and predicts an important 30-40 gain factor on the limiting fluxes after mitigation of the sky and guider-related backgrounds, and a better image quality (on a nominal acquisition time: 2h). The different subplots provide the remaining major offenders of the instrument. A simple click-change of the x-axis allows one to detect changes of parameters that most affect the final SNR and which enables grading correction and tolerancing. After these first-order analyses, the ETC allows one to further analyze/optimize other subtle tradeoffs[2]:

---

[1] The main interest of such devices is the ability, due to ~1000 stochastic amplifications of the incoming electrons and under certain conditions, to detect single photo-electrons. The amplification generates an excess noise factor that can be removed by post-processing (e.g.: thresholding photon-counting [5]). Unfortunately, this method is efficient under conditions that are difficult to reach at our temperature levels as read noise and smearing conspire against its efficient use.

[2] The full complexity of these tradeoffs can not be totally encompassed by an ETC as it would require adding a risk component, uncertainty, tolerancing, the preference for smooth low-risk optimums than higher but sharper SNR: for instance the loss of pointing if we add a shutter during a longer readout time, the interest of having data usable with and without thresholding, temperature-induced complexities such as condensation, the possible improvement of smearing inversion algorithms, etc.

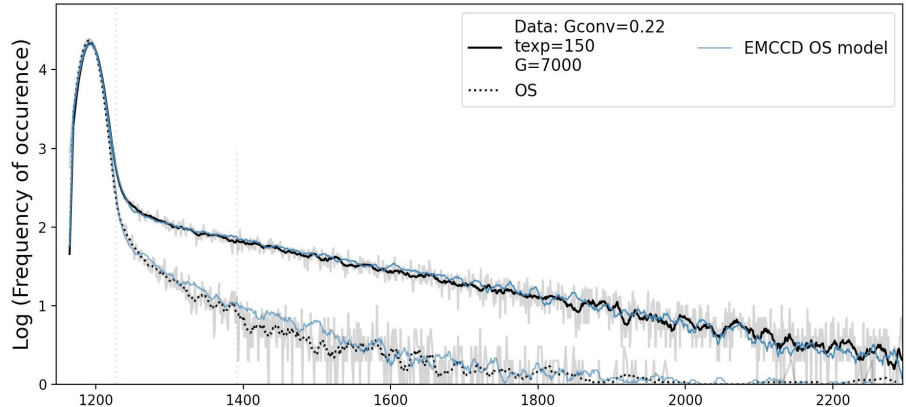

**Figure 3.** Unique electron spectrum. This fit method, based on the simultaneous fitting of the physical and overscan regions of the EMCCD (where parallel CIC and flux are not present) recovers most of the EMCCD parameters.

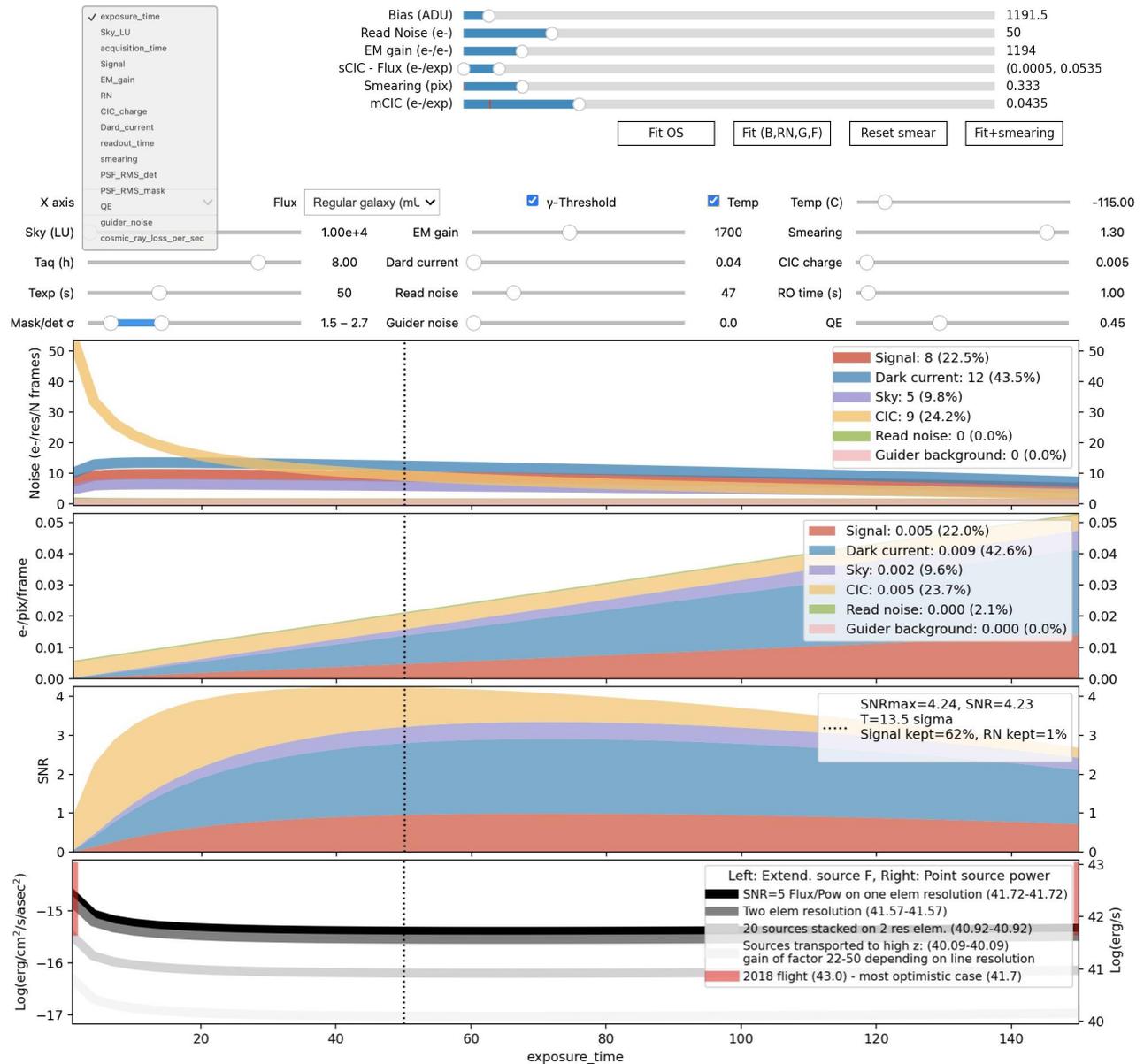

**Figure 4.** FIREBall-2 ETC and visualization front end GUI to visualize the evolution of the contributions/noise budget, SNR, and limiting flux with respect to any configuration of the mission/instrument/EMCCD parameters (see Section 4.1).

- **Exposure time:** this tradeoff (as shown in Fig. 4) results from an important read noise contribution at short exposure times, important surface loss due to the ~10/sec CR impact rate for long exposures, and the photon counting thresholding efficiency that decreases significantly when the number of electrons per pixel per exposure gets too high/low.
- **Detector temperature:** while dark current increases exponentially with temperature, charge transfer gets less and less efficient as the temperature decreases (below 190 K). The parameterization of the dark current and CTE temperature dependency allows one to derive the optimal temperature tradeoff.
- **Readout frequency**: the new version of the EMCCD controller enables reading out the image at 1 MHz instead of 10 MHz. Although this slower rate can decrease the read noise by a factor of 2 and might improve the CTE, it also increases the readout time by a factor of 5 to 10 (which might require the use of a shutter during image reading). At our temperature, the use of the 1 MHz readout appears to increase the overall SNR.
- **Photon counting threshold:** the threshold used to hamper the excess noise factor in EMCCDs (only possible under strict conditions) depends significantly on different parameters. While a low threshold will increase the false event rate, a high threshold will generate a significant loss of signal, hence removing the thresholding efficiency and interest. The optimum SNR(threshold) and its shape depend mostly on the effective amplification gain (significantly impacted by smearing) to read-noise ratio that needs to stay >>10 in order to be efficient but is also impacted by the input flux and the other noise sources.

**4.2.** FIREBall-2 upgrades[3]

### 4.2.1. Stray light mitigation strategy

With the duration of the flight, the extreme stray light contamination observed during FB2's first flight was the major offender of the instrument sensitivity. Solving these two issues by returning the sky level and acquisition time back to their nominal values (<0.1×Signal, 8h), improves the 5σ limiting flux by a factor >10. A major effort was undertaken into correcting this key issue.

By developing both sequential and non-sequential Zemax OpticStudio models of FB2, a comprehensive baffling system was designed [7], manufactured, and installed on the FB2 gondola and spectrograph (Fig 5):

- the gondola cap (b) will prevent the light reflected by the balloon to enter directly into the tank via the hole in the siderostat (the direct view of the moon was already baffled).
- the pupil plane baffle (large horizontal baffle on subfigure c) prevents off-axis light that might enter the medium and lower bench.
- the detector baffle (h) was designed to mitigate the scattered light incident on the backplane of FF2 which might reach the detector.
- the grating mask (g) obscures the unused portion of the diffraction grating.
- the tongue baffle on FF2 (e) blocks out-of-order diffracted light from reaching the detector and prevents light from entering through the FF2 hole before folding and focusing at SC2.
- the SC1 barrier was designed to prevent the specular reflection at the edge of the SC1 mirror chamber which might reach the detector.

### 4.2.2. Detector controller upgrade

The flight NüVü V2 CCCP controller has been replaced with the V3 upgraded version, which has the capability of reading out at 1 MHz (*vs.* 10 MHz in V2) and an improved clock resolution of 5 nanoseconds (*vs.* 10 nanoseconds). These two changes allow reducing the read noise by a factor of >2, which is critical when using these detectors as photon-counting devices. FB2 provides a key platform to test and demonstrate NASA Flagship-class technologies. Indeed, the V3 CCCP controller was developed for ROMAN, the Flagship-class NASA mission for the 2020s, and for the UV spectrograph onboard HabEx, one of four Flagship mission concepts presented for the Astro2020 Decadal review.

### 4.2.3. Focal-corrector realignment

FIREBall-2 focal corrector, which is a fast two-optics system magnifying and providing a large field to cover the slit masks, is a complex optics that was de-bonded at landing. Realignment has been the most challenging aspect for a re-flight because of the extreme impact on resolution and the low magnification tolerance (<1%).

---

[3] This list is not exhaustive and focuses on the most important aspects. Significant work has been done on other matters such as the implementation of a flight calibration box (with Zinc emission-line and Deuterium continuum UV lamps) or the optimization of the science masks to include more QSO and reduce slits overlapping

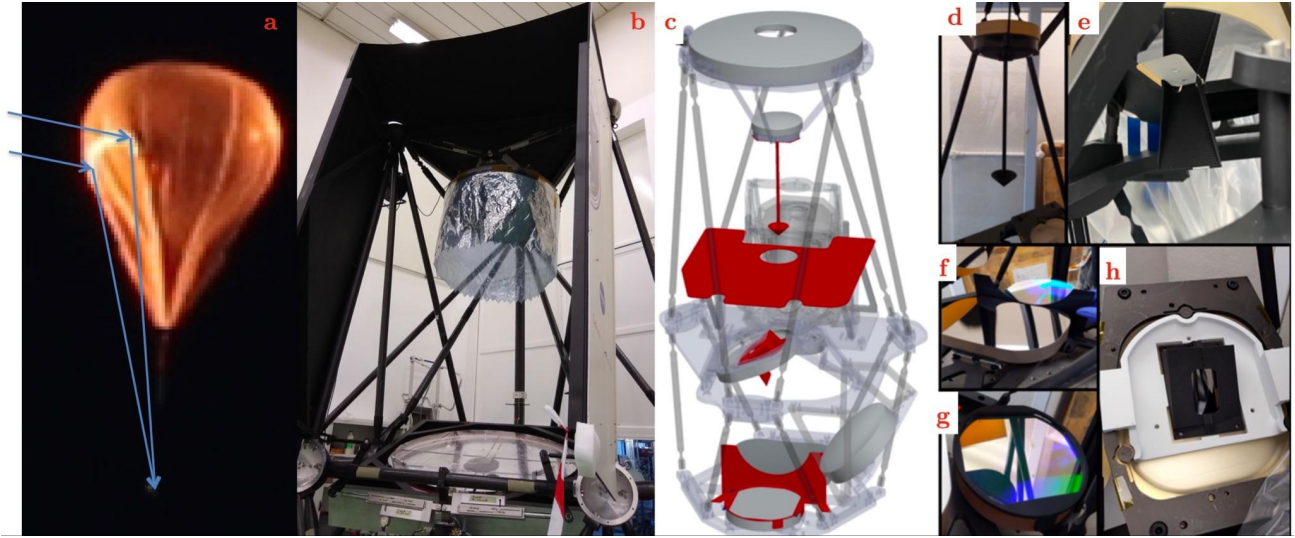

**Figure 5. a)** light path of the scattering light that goes through the siderostat hole after reflection on the deflated balloon. **b→e)** The different stages of the baffling system (see Section 4.2.1 for description). The conical baffle (d) was already implemented for the 2018 flight to suppress the double and triple bounce paths from reflecting through the focal corrector but, being in foil, was badly damaged during the hard landing. It has been replaced.

| INSTRUMENT | | | OPTICAL RESOLUTION / THROUGHPUT | | |
|---|---|---|---|---|---|
| **Parameter** | **2018 flight** | **Expected next flight** | **Parameter** | **2018 flight** | **Expected next flight** |
| FOV | 11' x 35' | ✓ | Spectrograph spat. | ~4'' | ~3'' |
| Band pass | 1970 - 2130 Å | ✓ | Spectrograph spect. | 1800 | 2100 |
| Total science acquisition | 30' | 2h | End-to-end spatial res. | ~8'' | ~5'' |
| Spacing between slits | ~5" | 20" | 80% encircled energy Ø | 10'' | ✓ |
| Red leak | ~100 x signal | ~0.1 x signal | Throughput | ~17% (11% with LO) | ✓ |
| **DETECTOR** | | | **GUIDING / POINTING** | | |
| Exp. time [Readout time] | 50s [1s] - 10MHz | 50s [5s] - 1MHz | Guider related background | >2xSignal (due to fiber comm) | ~0.1 x signal |
| Amplification | 1400 e⁻/e⁻ | 2000 e⁻/e⁻ | guider FOV | 15' x 15' | ✓ |
| Smearing exp. length - CTE* | 0.7pix - 80% | 0.3pix - 95% | Centroiding accuracy | 0.7" | ✓ |
| QE | ~50% | ✓ | | | |
| Dark current | 0.6 e⁻/pix/h | ✓ | gondola FOV [Az x El] | 360 x [40-64] | ✓ |
| CIC | 0.008 e⁻/pix | ✓ | Absolute pointing acc. | 3" over 4500s (3 σ) | ✓ |
| Read-noise | 107 e⁻/pix | 40 e⁻/pix | Pointing stability (1σ/axis) | 0.6" per 1000s | ✓ |

**Table 1.** Summary of the FIREBall 2018 flight performance and the expected performance for the next flight in 2023. The small horizontal CTE value is due to the extremely low temperature at which we have used the detector (150 K). On-ground dark current measurements were carried out prior to flight due to a guider-related light leak during the flight. The ✓ symbol in the "*Expected next flight*" column indicates that the required performance is already met. * The device temperature during the CTE measurements was 160 K in the 2018 flight, and around 180 K during the 2022 tests.

After rebonding, the damaged pad, one of the other two pads holding the FC secondary mirror debonded, one of the other two pads debonded, possibly due to stress or force since the 2018 flightshock (e.g. during shipping). All pads will be debonded in the coming months and the focal-corrector optics re-aligned.

### 4.2.4. Guider background mitigation

The communication between the guider camera and the guider computer is done through an infra-red fiber which leads to a red leak at the detector level. Even though it was mitigated by at least an order of magnitude thanks to a fiber sheath, a very low residual red leak of about an electron was observed. To further decrease this residual stray light leakage through the guider front window, an infrared blocking filter will be added in front of the window, along with a better baffle between the guider and the detector.

## 5. Conclusion

After its first flight in 2018, many upgrades have been carried out for FIREBall-2 (comprehensive baffling scheme, spectrograph, FC re-alignment, EMCCD controller upgrade, multi-slit target optimization). The ETC presented in this paper is a major new development that has been truly helpful for understanding the instrument tradeoffs, guiding some decisions, and designing some mitigation strategies. Over the next year, the remaining upgrades will ensure a flight in 2023 under nominal conditions.

FIREBall-2 is a complex suborbital payload and gondola due to the elaborated systems involved: state-of-the-art optics (focal corrector, grating, UV coatings), cutting-edge EMCCD used under extreme conditions, ice fusion and water boiling cooling techniques, and sub-arcsecond 4-axis guidance system. FB2's high potential in terms of scientific discovery as well as a technology demonstration and readiness level advancement comes with important risks. As for numerous projects, it has been significantly impacted by COVID, which coupled with the rapid turnover of Ph.D. students and postdoctoral researchers, makes the transfer of expertise more complex. In accordance with this, the goal for the next flight planned for 2023, was to solve/mitigate the different issues that appeared during the 2018 flight. Only minor changes were/are being performed to minimize the appearance of new risks and to ensure the full benefit of the factor 30 efficiency gain from the 2018 flight. The following flights will entail further upgrades, and allow FIREBall-2 to test new technologies such as:

- Use state-of-the-art visible blocking detector filters. The JPL multilayer metal-dielectric filters offer high in-band efficiency and out-of-band rejections by 3-4 orders of magnitude.
- Use new UV narrowband coatings to reject photons far from the instrument bandpass. FB2 was initially supposed to incorporate such coatings, but due to a long-distance scattering, they were not used. However, the huge improvements made over the past few years makes this ambitious endeavor feasible again.
- Use an upgraded telemetry system to increase the downlink maximum rate from to >10 Mbit/s (rather than the current 1 Mbit/s) enabling to download science images as well as critical housekeeping data.

## 6. Acknowledgments

FIREBall-2 is co-funded by CNES and NASA. The US funding comes primarily from the APRA program for sub-orbital missions. CNES and CNRS provided support for the French team in the FIREBall collaboration. Ground support was provided by the Columbia Scientific Balloon Facility (CSBF) during the different integration/flight campaigns. The JPL detector team gratefully acknowledges the collaborative effort with Teledyne-e2v.